\documentclass[sigconf]{acmart}
\AtBeginDocument{%
  \providecommand\BibTeX{{%
    \normalfont B\kern-0.5em{\scshape i\kern-0.25em b}\kern-0.8em\TeX}}}

\setcopyright{acmcopyright}
\copyrightyear{2021}
\acmYear{2021}
\acmDOI{10.1145/1122445.1122456}

\acmConference[WWW '21]{The Web Conference '21}{April 19--23, 2021}{Ljubljana, Slovenia}
\acmBooktitle{WWW '21,
  April 19--23, 2021, Ljubljana, Slovenia}
\acmPrice{15.00}
\acmISBN{978-1-4503-XXXX-X/18/06}

\usepackage{xcolor}

\begin{document}
\title{Controlling the Risk of Conversational Search via Reinforcement Learning}

\author{Zhenduo Wang}
\email{zhenduow@cs.utah.edu}
\orcid{1234-5678-9012}
\affiliation{%
  \institution{University of Utah}
  \city{Salt Lake City}
  \state{Utah}
  \country{United States}
  \postcode{84112}
}

\author{Qingyao Ai}
\email{aiqy@cs.utah.edu}
\affiliation{%
  \institution{University of Utah}
  \city{Salt Lake City}
  \state{Utah}
  \country{United States}
  }

\renewcommand{\shortauthors}{Trovato and Tobin, et al.}

\begin{abstract}
Users often formulate their search queries and questions with immature language without well-developed keywords and complete structures. Such queries are likely to fail to express their true information needs and raise ambiguity as fragmental language often yield various interpretations and aspects. This gives search engines a hard time processing and understanding the query, and eventually leads to unsatisfactory retrieval results. An alternative approach to direct answer while facing an ambiguous query is to proactively ask clarifying questions to the user. Recent years have seen many works and shared tasks from both NLP and IR community about identifying the need for asking clarifying question and methodology to generate them. An often neglected fact by these works is that although sometimes the need for clarifying questions is correctly recognized, the clarifying questions these system generate are still off-topic and dissatisfaction provoking to users and may just cause users to leave the conversation. 

In this work, we propose a risk-aware conversational search agent model to balance the risk of answering user's query and asking clarifying questions. The agent is fully aware that asking clarifying questions can potentially collect more information from user, but it will compare all the choices it has and evaluate the risks. Only after then, it will make decision between answering or asking. To demonstrate that our system is able to retrieve better answers, we conduct experiments on the MSDialog dataset which contains real-world customer service conversations from Microsoft products community. We also purpose a reinforcement learning strategy which allows us to train our model on the original dataset directly and saves us from any further data annotation efforts. Our experiment results show that our risk-aware conversational search agent is able to significantly outperform strong non-risk-aware baselines. 
\end{abstract}

\keywords{conversational search, reinforcement learning}


\settopmatter{printfolios=true}
\maketitle
\section{Introduction}

Besides the limitations of retrieval models and algorithms, the most important reason for the retrieval of unreliable results in modern IR systems is the vague information requests provided by users. Unsurprisingly, many users don’t have clear ideas about what they are looking for or how to effectively express their need to IR systems. This creates difficulties for the retrieval of high quality results. To 
address this problem, a diversity of techniques and UI design has been proposed to help users conduct effective search, such as query auto-complete \cite{autocomplete}, query suggestion \cite{querysuggestion1, querysuggestion2, querysuggestion3, querysuggestion4}, etc. In particular, conversational retrieval that actively refines user requests and search results by asking clarifying questions have been widely believed to be the key technique for future IR \cite{framework2017}. There are more and more studies and shared tasks \cite{whatdoyoumean, cqidentify, raodaume2019, xuasking2019, askingcq,zamania} \cite{trechard,aliannejadi2020convai3} on how to identify clarifying question needs and retrieve or generate good clarifying questions so that we could use the responses provided by users to prevent the retrieval of risky results. 

Existing studies on conversational retrieval, however, often ignores the possibility that conversational search/recommendation paradigms themselves could bring risk to users and modern IR systems. Previous work on this topic assumes that users are dedicated to search sessions after they submit a information request and would provide responses to any questions asked by the system \cite{aorb, systemaskuserrespond, conversationalrecommend, negativefeedback}. Such over-optimistic assumption neglects the effect of asking a bad question to users. For example, a user could be offended by a question with racial innuendo \cite{racial}; a user could be overwhelmed by over-specific questions; or a user could provide noisy responses to questions that have unclear intents, etc. All these cases would reduce user’s satisfaction, increase information accessing cost, and eventually drive users away from using the IR system again. 

\begin{figure}[t]
  \centering
  \includegraphics[width=\linewidth]{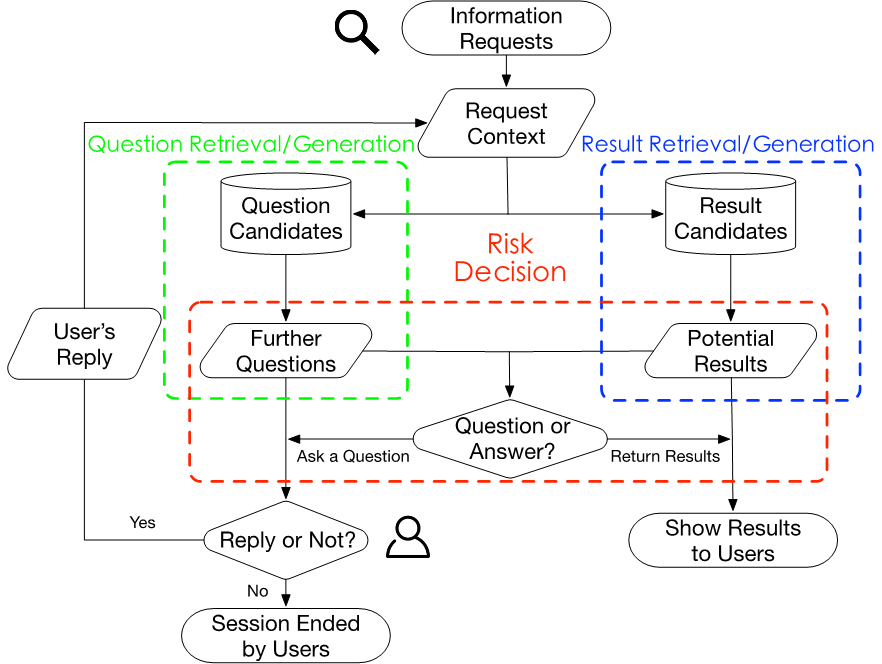}
  \caption{Risk aware conversational search agent}
  \Description{Risk aware conversational search agent}
  \label{riskawaremodel}
\end{figure}

In this paper, we propose to control and balance the risk and cost of result reliability and user engagements in modern IR systems with a risk-aware conversational retrieval paradigm. Formally, we believe that an effective and reliable conversational retrieval system should consist of three components as shown in Figure~\ref{riskawaremodel}: (1) A result retrieval module that retrieves and ranks candidate results according to the current information requests and corresponding context; (2) A question retrieval module, similar to those in previous studies \cite{msdialogrank, askingcq}, that finds or generates clarifying questions for users, hoping that user’s responses could enrich the context and further improve the quality and credibility of results provided by the result retrieval module; And (3) a risk decision module, unique in our proposed framework, that decides whether the system should ask the question provided by the question retrieval module or directly show the documents provided by the result retrieval module. The risk control module computes not just whether it should ask clarifying question or answer the query based on the context, but whether the answer or question retrieval model's response is better and which response has higher reward and lower risk. The risk decision module is the key for our proposed system to control and balance the risk of asking poor clarifying questions or providing immature results to users.

To demonstrate the effectiveness of the proposed conversational search agent, we build a conversational search agent with the above three components by combining state-of-the-art response retrievers and a decision making model trained by reinforcement learning. Through experiments we show that our risk aware conversational search agent is able to perform well on a conversational search dataset collected from Microsoft product community named MSDialog. And our agent is able to outperform baseline strategies such as directly answering the question or always asking 1 or 2 questions before answering the question. We consider our contribution as followings: (1) we propose a new risk-aware conversational agent which takes the risks into consideration when deciding to ask clarifying questions or answer the query given conversation context. The risk-control module can work as an add-on with any answer and question retrieval module.  (2) we propose a reinforcement learning approach to train proposed agent without having annotated data for when to ask clarifying question and when to give answer. (3) Through simulation experiments with different user models, we show that our risk-aware conversational search agent could improve both answer quality and user experience in interacting with the retrieval system. We make our project publicly available on GitHub.
 \footnote{https://github.com/zhenduow/conversationalQA}

\section{Related Works}
\subsubsection*{\textbf{Conversational search}}
The problem of the vagueness of user query has been studied by many previous works. One approach is to analyze the links among conversation turns. Kaiser et al. \cite{wordproximitynetwork} build a word proximity network to compute word coherence in addition to query-response relevance. Aliannejadi et al. in \cite{relevantturn} estimate relevance between current and previous turns and incorporate it in context representations. Another line is to alter or enhance the query by performing query suggestion or query auto-completion. The former provide user with several possible clearer queries that are close to user's query. Boldi \cite{boldi2008, boldi2009} uses query flow graph for query suggestion. Rosset et al. \cite{rosset2020leading} propose to suggest query by its usefulness. Query auto-completion emphasizes more on additive changes. Voskarides el al. in \cite{queryresolution} resolve anaphoras in later conversation turns by traversing through all previous turn terms and determine whether to add each of them to the current query which is modeled as a bag-of-words. 

Conversational search can usually be referred differently due to the complexity of conversation structures. Works such as \cite{yangnextquestion, topicpropagation, qu2020open} study conversations in a session where user can ask multiple questions potentially with topic drift. We consider this as fundamentally different from the conversations we are interested in, where the whole conversation is centered around the user's initial information need.

\subsubsection*{\textbf{Asking clarification questions}} A considerable amount of attention has been put on asking clarification question by natural language processing and information retrieval community. Earlier in the TREC 2004 HARD track\cite{trechard}, asking clarifications is made available for participants to get additional information. Rao and Daumé III in \cite{raodaume2018} propose a clarification question selection model by maximizing the expected value of perfect information after getting the answer to the clarification question. Later in \cite{raodaume2019}, they extend the model to a reinforcement learning model which better fits multi-turn settings. Recent works approach asking clarification questions in various ways. Aliannejadi et al. \cite{askingcq} select from human generated questions and Zamani et al. create a clarifying question taxonomy based on user studies and \cite{zamania} solve automatically generation of clarification questions via multiple approaches. Later, Zamani et al. \cite{zamani2020mimics} also publish an annotated dataset consisting of real-world user interactions. Cho et al. \cite{askingcqnlgmulti} generate one question that can capture common concepts in multiple documents. As why and how to retrieve or generate clarification questions have been studied by all these previous works, some recent works also study when to ask clarification question, i.e., identifying the need. Xu et al. \cite{xuasking2019} create a clarification identification and generation dataset where they solve clarification identification as a binary classification problem. Most recently, Aliannejadi et al. \cite{aliannejadi2020convai3} organize a shared task which pose questions on when to ask clarifying questions and how to generate them, concluding all the aforementioned works. 

\subsubsection*{\textbf{Risk control}}
Radlinski and Craswell in \cite{framework2017} propose a theoretical framework for conversational search, where they highlight the nature of multi-turn interactions between 2 participants (user and agent) in conversational search problem. They also define action spaces for both the user and the agent in their work and emphasize the necessity of a model which can make decisions among the actions. In our work, we implement a similar decision making model for the agent but we simplify the action space of the agent into either ask clarification questions or provide retrieval results as answer. And we use the ground truth response in the dataset to simulate user instead of modeling the user. 

Few existing works have address the conversational search risks in making these decisions, which is a key factor in real-world IR applications. Su et al. \cite{surisk} propose a risk control framework for web question answering where their main reading comprehension model is followed by a qualify model which estimates the predicative uncertainty and a decision model which makes decisions of whether or not to answer the question depending on the qualify model's output. Our system extends their decision model by allowing the agent to ask clarification questions instead of doing nothing when the query is not answerable.

\subsubsection*{\textbf{Dataset}}
In \cite{msdialogintent, msdialogrank, msdialogintent2}, Qu and Liu introduce a new dataset MSDialog that consists of conversational question answering thread collected from Microsoft forum. A large amount of questions fall into one of the major Microsoft products such as Windows, Office, and Bing. In their works, they preform two tasks on the dataset. Qu analyze and characterize user intent of each utterance in the conversation, in which they include clarification question as one type of intent. Liu use the conversation to train a next-response-prediction reranker. Our task operates on the whole conversations via simulating the conversation process between user and agent.

\section{Interactive User-Agent Model}
Our risk-aware conversation agent aims to understand and answer user's query by iteratively interacting with user model in multiple turns. Starting from user's initial query as the first turn, in each turn, the rerankers first rank the answer and clarifying question. Then a decision maker uses the ranking results to decide whether to give answer to the query or ask clarifying question. If the decision is to ask question, then the system would show the clarifying question to the user and collect their feedback, which may give the agent more information to retrieve better results in the next turn.




\subsection{Answer and Question Reranker}
The goal of answer and question reranker is to retrieve the best answer or clarifying questions from candidate pools based on the user's query and the current context information. To demonstrate that our decision maker can work with any answer and clarifying question retriever/reranker model, we test different rerankers with our decision maker. The rerankers we test are the ParlAI bi-encoder and poly-encoder rerankers \cite{polyencoder}. In each experiment, we use the same reranker structure for both answer reranking and clarifying question reranking but with two sets of parameters. 

\subsubsection*{\textbf{Bi-encoder}}The bi-encoder structure uses two transformers \cite{bert} encoders $E_Q$ and $E_P$ which are pretrained on Wikipedia and Toronto Books to encode the conversation context $q$ and candidates $P_k =\{p_1, ..., p_k\}$. Candidates can either be answers or clarifying questions, depending on the ranking task. Since the context and candidates are independently encoded, the segmentation token are both 0. A special token [S] is append to the start of context sentence and the representation for this special token is chosen as the representation for the whole sentence, same for candidate sentences. After encoding the texts into vectors, the bi-encoder scores each candidate by computing the dot product between the candidate encoding vector and the context encoding vector. Then all the candidate are ranked using the scores. The computation of the score can be represented as follows:

\begin{equation}
s_p = E_P(p)^\top E_Q(q)
\end{equation}

\subsubsection*{\textbf{Poly-encoder}}The poly-encoder uses separate encoders just like bi-encoder. It first encodes the context $q$ and the answer or clarifying question candidates $P_k =\{p_1, ..., p_k\}$ with pretrained transformers. But unlike the bi-encoder where there is only self-attentions in context and candidates, poly-encoder allows con-attentions between context and candidates. The context $q$ can attend to multiple learnable codes $(C_1, ..., C_m)$ and generate multiple attended context vectors $(q_{\text{attn}}^1, ..., q_{\text{attn}}^m)$:
\begin{equation}
q_{\text{attn}}^i = \sum_j w_j^{C_i}h_j 
\end{equation}
where $(w_1^{C_i},..., w_N^{C_i}) = \text{softmax}(C_i\cdot h_1, ..., C_i\cdot h_N)$ and $(h_1, ..., h_N)$ are encoder outputs.

The poly-encoder then computes the encoded candidate vector $E_P(p)$ just like in bi-encoder. With multiple context vectors and the candidate vector, poly-encoder computes a candidate-attended context vector $q_{\text{attn}}$:
\begin{equation}
q_{\text{attn}} = \sum_i w_iq_{\text{attn}}^i
\end{equation}
where $(w_1,..., w_m) = \text{softmax}(E_P(p)\cdot q_{\text{attn}}^1, ..., E_P(p)\cdot q_{\text{attn}}^m)$.

Then the reranking score is computed as:
\begin{equation}
s_p = q_{\text{attn}}^\top E_P(p)
\end{equation}

The bi-encoder can be seen as computing the similarity between context and candidates on the sentence-level without inter-attentions between the context and candidates. This makes the structure of bi-encoder simple and running time shorter. A straight forward addition of inter-attention is to concatenate context and candidates together (referred as cross-encoder in \cite{polyencoder}). This model first concatenate the context text and the candidate text together, and append a special [S] token to the beginning, just like bi-encoder. Then it uses a pretrained transformer encoder $E_{QP}$ to encode the concatenated text and use the representation of token [S] as the representation for the concatenated text. The score of the candidate is finally computed as:

\begin{equation}
s_p = E_{QP}(q,p)^\top W
\end{equation}

where $W$ is a learnable projection matrix to reduce the encoding vector to a scalar score. 

This cross-encoder structure can improve the performance but will increase running time by hundreds of times, rendering it not deployable in real world applications. Hence we do not include it in our experiments. The poly-encoder structure simplifies the necessary but expensive inter-attention computations and gets the best of both bi-encoder and cross-encoder. It is able to achieve nearly the same performance as cross-encoder in experiments.

\subsection{Decision Making Model}
The decision maker uses a deep Q network (DQN) to decide between answering the query with the best answer and asking a clarification question with the best question. The DQN first uses a BERT-based encoder to encode the initial query $q$, context history $h$, the top $k$ clarifying question $\{cq_1,...,cq_k\}$, and the top $k$ anwer $\{a_1,...,a_k\}$ and take the [CLS] token vectors as their feature representations. Then it reads the reranking scores $s_{cq}^{1:k}$ and $s_{ans}^{1:k}$ of the top $k$ questions and answers from the reranker output. Finally, it concatenates the all the features and uses a 2-layer feedforward neural network to generate a $2\times 1$ decision vector $y_{\text{pred}} = (r_{ans}, r_{cq})$ of the predicted rewards for giving the answer and asking the clarifying question. The decision maker DQN can be seen as an essentially binary classification network and represented as:

\begin{equation}
D(S) = W_2\cdot \phi(W_1\cdot\theta + b_1) + b_2
\end{equation}

where state vector $S = (q, h, cq^1,..., cq^k, a^1,...,a^k, s_{cq}^{1:k}, s_{ans}^{1:k})$ is the concatenation of all the features, and $\phi$ is ReLU activation function. Note that our decision maker DQN does not have activation layer in the second layer, since it predict the action rewards (logits) instead of action classification probabilities.

\begin{figure}[h]
  \centering
  \includegraphics[width=\linewidth]{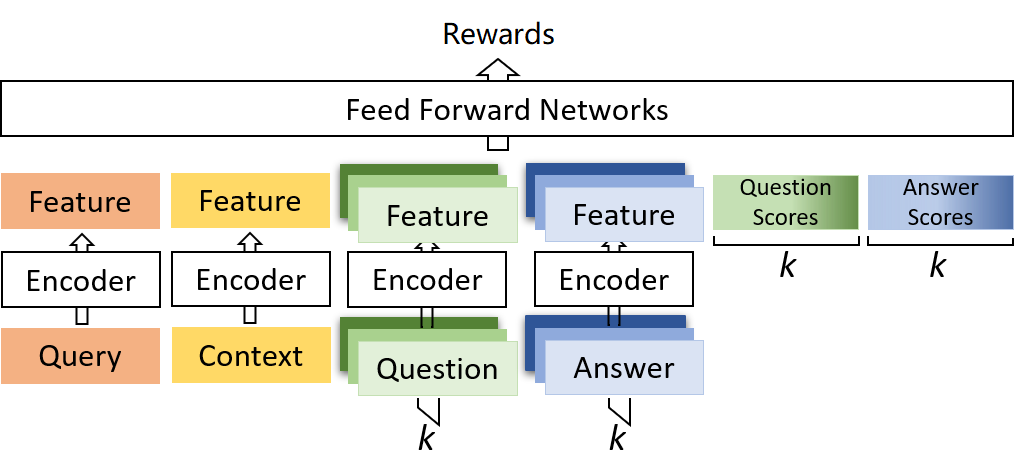}
  \caption{Decision Maker Deep Q Network}
  \Description{Decision Maker Deep Q Network}
  \label{DQN}
\end{figure}

The structure of the DQN is shown in Fig~\ref{DQN}. Although the structure of DQN is simple, we do not have annotated data to supervise the training of this network. To overcome this challenge, we use reinforcement learning, which trains the network by simulating the interaction between agent and user. We define reward and penalty for each interaction outcome using simple settings. The DQN is trained by learning from these reward or penalty. We will describe this in later section.

\subsection{Training}


\subsubsection*{\textbf{Rerankers}}For our answer and question rerankers, we download the checkpoint that was pretrained on huge Reddit dataset \cite{redditdataset} as the beginning points for tuning. Then we fine-tune it on the answer reranking and question reranking datasets to train the answer reranker and question reranker separately. The reranker model is trained to minimize the average cross-entropy loss between the batch relevance label vector and the batch similarity dot product logits vector. During tuning, we divide the training set into batches with 100 conversation-response pairs and use the true response as positive sample and all other responses in the batch as negative samples. 

\subsubsection*{\textbf{Decision Maker}}After pretraining the rerankers, we fix them during the training of DQN. We use reinforcement learning to train our DQN without supervision. At the beginning of training, the DQN randomly makes decisions between answering (A) the query and ask clarification questions (CQ). Each decision will result in an reward (or penalty). Then it learns to make decisions from the reward $r$ of its action $a$. We emperically define action rewards as follows. If the agent chooses to answers the question, the conversation immediately ends and it receives reward equal to the answer reranker's Mean Reciprocal Rank (MRR) of the reranked top $k$ answers. Hence a correct top answer will get full reward $r=1$, and a non-top-ranked correct answer can get partial reward in $(0,1)$. If the agent chooses to ask a question and the question turns out to be a good one, it receives a reward $r_{cq}$. Then the user simulator will respond with the true answer from the dataset to the question. The agent then updates the conversation context history with the question and the answer, and continue the conversation with user. If the agent chooses to ask a question but the question is bad, then it receives penalty $p_{cq}$ and is forced to end the conversation. We summarize this in the policy table in Table~\ref{DQN Policy table}.

Now we explain the intuition of our reward settings. The reinforcement learning algorithm trains the DQN to decide between answering or asking clarifying question by the immediate reward and potential future reward or risk. Giving an answer guarantees a non-negative answer reward, while asking question sacrifices the immediate reward for a potential higher answer reward in the future but is risky. 

\begin{table}[htbb]
\caption{Decision Maker DQN Policy Table}
\begin{center}
\begin{tabular}{|c|c|c|}
\hline
& Relevant & Irrelevant \\\hline
Answer & \multicolumn{2}{c|}{Answer Reciprocal Rank} \\\hline
Ask & $r_{cq}$ & $p_{cq}$ \\\hline
\end{tabular}
\label{DQN Policy table}
\end{center}
\end{table}

The goal of the decision maker DQN training is to train the DQN to predict the reward of an action $a$ (answer or ask clarifying question) given its input state $S$. The training starts by DQN randomly exploring an action given state $S$, then learn from all state-action-reward ($S$-$a$-$r_a$) tuples seen this way. Assume the action is to answer ($a=ans$), the predicted reward is: \begin{equation}
y_{\text{pred}}(S)_{ans} = DQN(S)_{ans}
\end{equation}
The true reward $r$ is computed as the answer's reciprocal rank:

\begin{equation}
y_{\text{target}}(S)_{ans} = r_{ans} 
\end{equation}
In this case, since the action is to answer, the conversation will always end. Alternatively, if the action is to ask clarifying question, and if the clarifying question is relevant, then:
\begin{equation}
y_{\text{target}}(S)_{cq} = r_{cq} + \sigma \cdot \max_{a^* = ans, cq} DQN(S^*)_{a^*}
\end{equation}
where $r_{cq}$ is the immediate reward for asking relevant question, $\sigma$ is the discount factor for future reward, $S^*, a^*$ are the updated state and action based on current state $S$ and action $a$. The second term is essentially a discount factor times the higher predicted reward between $ans$ and $cq$ after updating the conversation state $S$ to $S^*$. 
If the action is to ask but the clarifying question is found to be irrelevant, then:
\begin{equation}
y_{\text{target}}(S)_{cq} = p_{cq}
\end{equation}
where $p_{cq}$ is the penalty for irrelevant clarifying question.

Finally, the training goal is to minimize the mean squared error loss between $y_{\text{pred}}(S)_{a}$ and target reward $y_{\text{target}}(S)_{a}$ defined as: 

\begin{equation}
L = \text{MSE}(y_{\text{target}}(S)_{a}, y_{\text{pred}}(S)_{a})
\end{equation}

The DQN training starts with random exploration, but as the training progresses, we gradually allow the DQN to make decision based on its own prediction and learn from them. We train the DQN until it converges.
 
Reinforcement network is notorious for not guaranteed to converge during training. Hence we use experience replay strategy. Experience replay is an often used training strategy which make full use of previous seen samples and make the training more likely to converge. In experience replay, we also increase the experience play times of action-reward pairs of asking clarifying question. This can also be seen as data oversampling. We find that doing this can help the agent to learn better about the risk in asking questions. We only train our DQN with actions that result in non-zero rewards. This include correctly choosing good answer or question, and wrongly choosing to ask. Because we want our system to learn the risk of asking irrelevant clarification questions.

\subsection{Inference}
In the inference phase, the conversational search agent starts with the initial query from the user. It first ranks all answer candidates and clarifying question candidates. In each step, we sample 99 negative answers along with the ground truth answer to make an answer candidate set, same for the clarifying question candidate set. It then pass the query, the context, the top $k$ relevant answers, and the top $k$ questions to the decision maker, which decides whether to answer the question or to ask clarifying question. If answer action is chosen, then the conversation ends and we evaluate the answer using the Mean Reciprocal Rank. If asking clarifying question action is chosen and the question is good (is the true clarifying question), then we update the context with additional information and repeat the ranking and decision making process. If asking clarifying question action is chosen but the question is bad (is a negative sampled question), then we end the conversation. We will describe how we evaluate inference results in evaluation section.

\section{Experiments}

\subsection{Dataset}
We use the MSDialog dataset \cite{msdialogintent, msdialogrank, msdialogintent2} in our experiments. This dataset consists of question answering conversations from online forum Microsoft products. We process and filter the complete MSDialog dataset (MSComplete) with several criteria that we believe are necessary for the conversational search scenario. 

First, we need conversations between only two roles, namely the user and the agents. Because we suppose that more participants will lead to more topic divergence, and two-participant-conversation can best fit the information need/clarification scenario. In the MSDialog data, there are conversations which have multiple agents. We assume all agents share the same goal to help the user, and we merge these agents into one 'agent' role so that all the conversations have only 2 participants, namely the user and the system agent. We also merge consecutive responses from a participant into a single response so that all conversations are conducted in turns. 

After above preprocessing, we do a further filtering step and only keep conversations which are between 4 to 10 turns (so that at least one clarification question is asked and a few outliers with too many turns are left out) and also have a voted final answer (In MSDialog dataset, each conversation turn has a binary label indicating if this reply is voted as the final answer by community). The processing and filtering leave us a subset of 3,792 conversations. We also remove all conversation turns happened after the voted answer, thus we can focus on the process of answering the user question.

The above preprocessing and filtering steps result in a dataset with 3762 conversations. For the rest of conversations that are either missing a correct answer or being too long, we set them aside and use them for training of the question rerankers. The set aside data has 19,793 processed conversations. Our dataset statistics can be found in Table~\ref{datasetstat}.

\begin{table}[htbb]
\caption{MSDialog and our dataset statistics}
\begin{center}
\begin{tabular}{|l|l|l|}
\hline
item             & MSDialog & Ours  \\
\hline
\# conversations & 35,000 & 3,762  \\
\hline
Max. turns        & 1700 & 10      \\
\hline
Min. turns        & 3 & 4       \\
\hline
Avg. turns        & 8.94 & 4.70   \\
\hline
Avg. \# words per utterance  & 75.91 & 65.16   \\
\hline

\end{tabular}
\label{datasetstat}
\end{center}
\end{table}

\subsection{Experiment settings}
We design our experiments to test whether our risk-aware conversational agent can find good answer for user's query as well as improving user's experience. Given the user's initial query as input, the agent will iteratively interact with the user in a multi-turn conversation. In each turn, the agent will choose response in the following three steps:
\begin{enumerate}
    \item Rerank an answer candidate set based on the initial query and conversation history as context. The answer candidate set has 1 ground truth answer and $k$ randomly sampled negative answers. Suppose a multi-turn conversation can be represented as $\{u_1,a_1,u_2,a_2, ...,u_N,a_N\}$, where $u_1$ is the user's first utterance (i.e., initial query), and $a_1$ is the agent's first utterance and so on. The ground truth answer is always the agent's last utterance ($a_N$) for any turn $1,...,N$. (We truncate conversations so that the answer is always the last utterance). The negative examples are randomly sampled answers from other conversations in the dataset. 
    \item Rerank a clarifying question candidate set based on the query and context. The clarifying question candidate set has 1 ground truth clarifying question and $k$ randomly sampled negative questions. The ground truth clarifying question for a turn $i, i=1,2,...,k-1$ is the $i$th utterance of agent ($a_i$). In the last turn, there is no correct clarifying question for the agent because $a_N$ is an answer not a question. The negative examples are randomly sampled clarifying questions from other conversations in the dataset. 
    \item Use risk-aware decision maker to decide between providing answer or asking a clarifying question using the top results from (1) and (2). Finally, we use the answer/clarifying question favored by the decision maker as the response of the current turn. 
\end{enumerate}

After giving response. our agent will wait for user model's response and either enter the next turn or stop and receive a final score depending on user's response. We build a user simulator model to respond to agent based on the following rules. 

\begin{enumerate}
    \item If agent chooses to provide answer, the conversation will end. User then awards agent a final score which is computed as the provided answer's ranking metric. 
    
    \item If agent chooses to ask a clarifying question, then the user checks if the clarifying question is appropriate (if it is in the original conversation). This is essentially checking if agent's response is in $\{a_1,a_2,...a_{N-1}\}$. If yes, then user simulator will respond with user's follow-up answer in the data. Otherwise, the user will mark the question as off-topic. As mentioned in \cite{zamania}, users are not necessarily angry seeing bad clarifying questions. Hence we assume that there are various types of users in real-world, hence we test different user models during experiments. The main difference of these user models is how many times they can tolerate off-topic questions. A 0-tolerance user simulator will immediately leave the conversation and also give the agent 0 score for asking bad clarifying question. And a 1-tolerance simulator will give the agent another chance for asking clarifying question by looking at the second best clarifying question from the rank, and then leave when seeing a bad clarifying question for the second time, etc.
\end{enumerate}
 
If the conversation continues, the agent and the user will repeat the above process. Otherwise, the agent is guaranteed to get a score that is either a ranking score of the provided answer or 0 for not able to give any answer to the user. Please refer to Figure~\ref{riskawaremodel} as our model flowchart.

\subsection{Baselines}
To demonstrate that our conversational search agent is able to produce better answers and reduce the number of bad questions, we conduct experiment and compare our agent with previous strong model and multiple baselines: (1) \textbf{Q0A}, a baseline that always directly answers the query given the initial query. (2) \textbf{Q1A}, a baseline that always asks exactly 1 clarification question and then answers the query. (3) \textbf{Q2A}, a baseline that always asks exactly 2 clarification question and then answers the query. (4) \textbf{CtxPred}, a risk-unaware baseline which first employs a binary classifier to predict whether to answer the query or ask clarifying question given current context. Then, according to the classification result, use answer or question reranker to rank and select the top response. This is a common approach by many previous works like \cite{xuasking2019,aliannejadi2020convai3}. (5) We also include an oracle model which always knows when to answer and when to ask clarifying question, given the specific answer and question reranker results. In each comparison experiment, we use the same reranker models and parameters for our model and baseline (1)-(5).

\subsection{Evaluation metrics}
We mainly use recall@1, Mean Reciprocal Rank(MRR) and the decision making accuracy to show our system and the baseline systems' strengths and weaknesses. Each of the metrics are different and measure the answer quality, model performance and user's experience from a unique aspect. Recall and MRR are already widely used in ranking in general. But since our evaluation is done on the result of entire conversation and could be different from generic ranking, we will explain the exact definition of these metrics in our evaluations.

\subsubsection*{\textbf{Recall@1}} As described in the model overview section, the agent can ask as many clarifying questions as long as those question are relevant to current conversation and do not make the user leave, but it only gets one chance to answer the question. The recall@1 metric is defined as the frequency of conversations where the user eventually get the ground truth answer regardless of how many clarifying questions are asked. If during the conversation, the agent asks a bad question and cause the user to leave, and hence the agent never have a chance to answer the user again, this conversation will get 0 as recall@1. We consider recall@1 as a direct measure for answer quality and closely related to user experience.

\subsubsection*{\textbf{Mean Reciprocal Rank}} Similar as the recall score, the Mean Reciprocal Rank (MRR) score also cares only about the final answer. In our experiments, we sample 99 negative responses with 1 positive response. So, the minimum MRR for answer is 0.1 by definition. But if agent fails to ask a good question before user runs out of patience and causes user to leave without getting any answer. Then the reciprocal rank score for this conversation will be 0. This means MRR severely punishes the model of asking bad clarifying questions. In order to get high MRR score, the agent must try its best to avoid asking bad clarifying question. Using MRR as primary evaluation metric is unfair since it favors conservative models which seldom ask clarifying question.

In real world application, only the top answer will be return to user, thus MRR cannot directly measure user experience. Because of this, we find that the MRR describe models' performances from an introspection aspect rather than user experience.

\subsubsection*{\textbf{Decision error}} The decision error measures the frequency of our agent making a worse decision between answering and asking question. We define `worse' as (1) asking a bad clarifying question or (2) answering the query when the answer is bad and asking clarifying question is better. According to our definition, worse decisions will always result in lower answer quality or cause user to leave. But non-worse decisions are not always the better, because asking clarifying question and having more information from the user does not necessarily improve answer quality in later turns. 

At the first glance, a non-worse decision will always get higher or at least the same MRR points by definitions. Yes, but this is only true for an individual case. In general, average lower decision error does not necessarily mean higher MRR. The reason is that, for example, when the agent correctly decides to ask clarifying question for more information, the final answer MRR may just improve from 0.2 (the 5th) to 0.33 (the 3rd). But when the agent makes a wrong decision by asking a bad clarifying question and cause the user to leave, the MRR will drop to 0. This implies that a few bad decisions can offset the MRR gain from many good ones. Compared with recall and MRR, decision error mainly measures models' ability of avoiding risks. Together with recall, decision error is closely related to user experience.

\subsection{Technical Details}

We use the implementation of bi-encoder and poly-encoder from ParlAI \footnote{https://github.com/facebookresearch/ParlAI/tree/master/projects/polyencoder} with modifications to corroborate our experiments. We implement our agent and user simulator from scratch based on Pytorch. We split our dataset into 5 folds and use cross validation to test significance. We run our main experiments on a single core of GeForce RTX 2080 Ti with 11GB memory. The pretraining of the bi-encoder and poly-encoder rerankers are run on 4 cores with a smaller batch size than the original settings of ParlAI's original settings.  

Through experiments, we test and tune our model hyper-parameters. We finally set the clarifying question reward $r_{cq}=0.21$ and penalty $p_{cq}=-0.79$, the future reward weight in reinforcement learning $\sigma = 0.79$. We train our model with a learning rate $lr=10^{-4}$ and regularization weight $\lambda = 10^{-2}$ for decision making DQN.

\section{Results and Analyses}

\begin{table*}[htbb]
\caption{Comparison of all models and baselines using poly-encoder as reranker. Numbers in bold mean the result is the best excluding oracle. $\ddag$ indicates $p < 0.01$ statistical significance over the best among baseline models.}
\begin{tabular}{l|l|l|l|l|l|l|l|l|l}
\hline\hline
Users   & \multicolumn{3}{c|}{0-tolerance} & \multicolumn{3}{c|}{1-tolerance} & \multicolumn{3}{c}{2-tolerance} \\ \hline
Models  & R@1/100     & MRR    & Dec. err    & R@1/100     & MRR    & Dec. err    & R@1/100     & MRR    & Dec. err  \\ \hline
Q0A     & 0.7475    & \textbf{0.8398}    & 0.1975    &0.7475    & 0.8398    & 0.1975  &  0.7475    & 0.8398    & 0.1975    \\ 
Q1A     & 0.7525    & 0.8044    & 0.1225    & 0.7850    & 0.8494   &  0.0650  & 0.8200    & 0.8723    & 0.0225   \\ 
Q2A     & 0.0075    & 0.0075    & 0.9925    & 0.0075    & 0.0075    & 0.9925  & 0.0125    & 0.0125    & 0.9875    \\ 
CtxPred & 0.7400    & 0.7960    & 0.1275   & 0.7850   & 0.8494   & 0.0600  & 0.8200    & 0.8723    & 0.0225            \\ \hline
Ours    & \textbf{0.7775}    & 0.8305   & \textbf{0.0975}$^{\ddag}$ &  \textbf{0.7875} &  \textbf{0.8530} &   \textbf{0.0575}$^{\ddag}$ & \textbf{0.8200}$^{\ddag}$  & \textbf{0.8723}$^{\ddag}$  &  \textbf{0.0225}$^{\ddag}$           \\ \hline
Oracle  & 0.8575    & 0.9139    & 0  &   0.8650  &   0.9169     & 0    & 0.8925    & 0.9324  & 0   \\ \hline\hline
\end{tabular}
\label{polyexperiment}
\end{table*}

\begin{figure}[t]
  \centering
  \includegraphics[width=\linewidth]{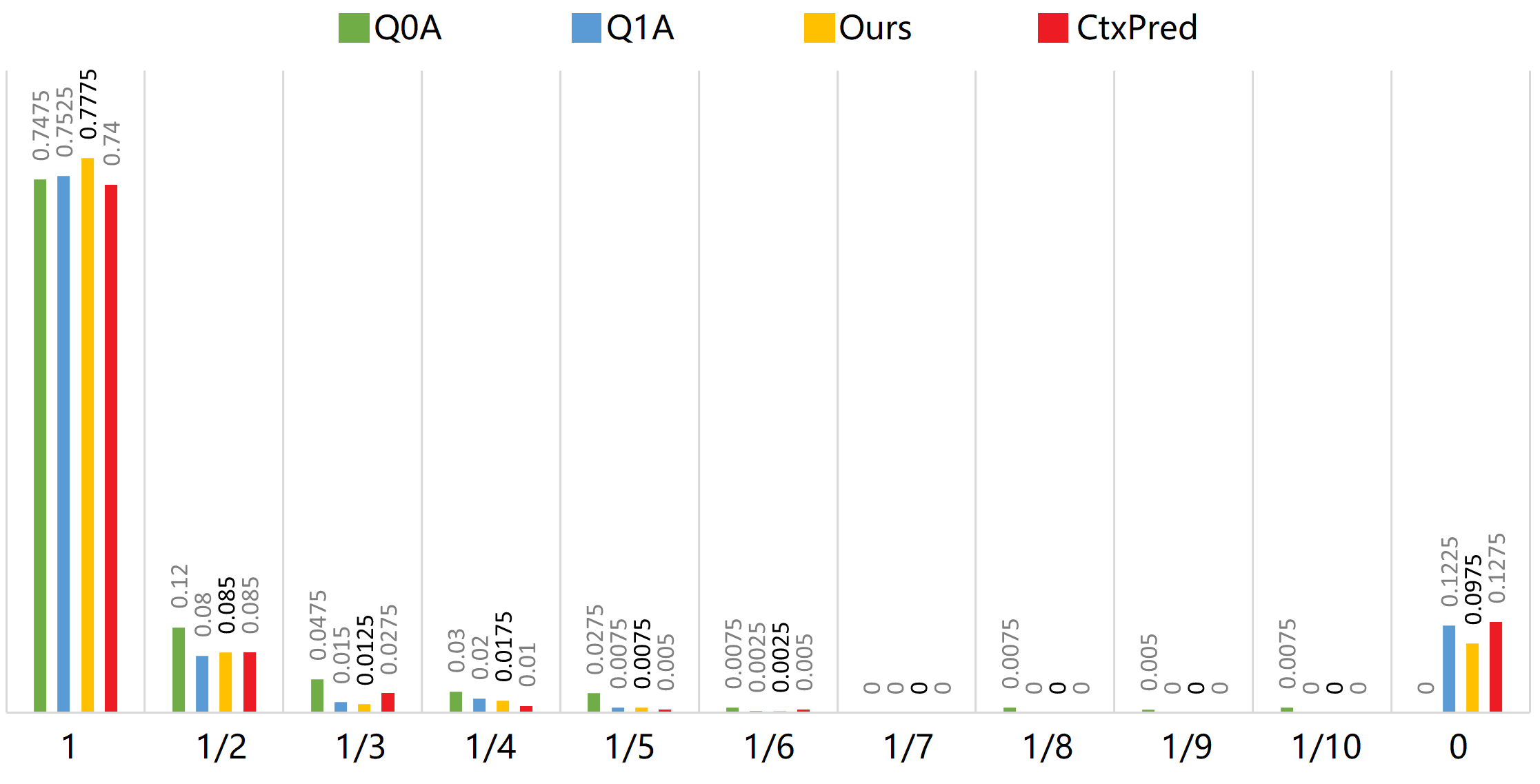}
  \caption{MRR Distributions for poly-reranker experiment and 0 tolerance.}
  \Description{MRR Distributions for poly-reranker experiment and 0 tolerance.}
  \label{distribution}
\end{figure}

To show our risk-aware conversational agent is able to give better answers than the baselines, we conduct multiple simulation experiments by interacting with different user models and combing our agent with different rerankers. In each experiment, we compare the all the baselines and oracle with our model when interacting with different user tolerance for bad clarifying questions. The two sets of experiment results are shown in Table~\ref{polyexperiment} and Table~\ref{biexperiment}. We abbreviate recall from top 1 in 100 candidates to R@1/100 and decision error frequency to Dec. err in our tables. We also show a comparison of MRR distribution of our model, Q0A, Q1A, and CtxPred baselines in Figure~\ref{distribution} of the poly-encoder experiment when user tolerance set to 0.

Our first experiment compare different user models when all of them use the poly-encoder reranker in Table~\ref{polyexperiment}. The most demanding user cannot tolerate any bad clarifying question that is irrelevant its information need, then the most tolerant user can tolerate up to 2 bad questions in total before leaving the conversation. For each user tolerance level, we compare the 6 baselines we mentioned in section 4.3. First of all, we can see from Figure~\ref{distribution}, all the models that can ask clarifying questions (Q1A, CtxPred, and ours) have 0 reciprocal rank score between $1/7$ and $1/10$, and their reciprocal score frequency between $1/2$ and $1/6$ are all lower than Q0A which never asks clarifying questions. These frequencies go to either the `1' bar or `0' bar. This implies that asking clarifying question can in general improve answer quality but also has the risk of reduce user experiences. Another observation from the tables is that the Q2A baseline has extremely low recall and MRR and high decision error. This is because that the conversations in our data have 4.7 turns on average, which means that the agent usually needs to ask only one clarifying question to give the answer to users' queries. Another reason is that for Q2A to get a correct answer, it has to be correct on three reranking tasks (2 question 1 answer), which is exponentially harder than other baselines. In fact, all the model except for Q0A baseline have this exponential error issue more or less.

From the tables we can see that our risk-aware conversational agent is able to outperform all the baseline methods in terms of recall@1 and decision accuracy when interacting with all types of users. This means in general, our risk-aware agent is able to answer more query correctly than all the baselines as well as improve user experience. We can also see that when user tolerance for bad clarifying question is 0, our agent does not score the highest MRR although it has the highest decision accuracy. We previously mentioned the possibility of this result in section 4.4. The reason for this can also be found in Figure~\ref{distribution}. the yellow bar stands for our model and the green bar is the Q0A baseline. The Q0A baseline by definition always answer the query immediately and thus will never ask a bad clarifying question which gets 0 reciprocal rank score. Although our agent is aware of the risks, it cannot completely avoid asking bad clarifying questions occasionally. Thus it gets 0 reciprocal rank score for $9.75\%$ of the time when it could get some answer score by just answering the query. This loss from asking bad clarifying questions offsets the gain of being able to give better answers. But if we compare the decision error of our model with Q1A and CtxPred baselines, we can see that our model still manage to get both the highest answer accuracy ($0.7775>0.7525>0.74$) and the least number of times asking bad clarifying question ($0.0975<0.1225<0.1275$). Combining all these observations above, we can conclude that asking clarifying questions can help improving answer quality in general but is very risky when facing a demanding user. And among all the non-trivial models, our risk-aware agent perform the best.

\begin{table*}[htbb]
\caption{Comparison of all models and baselines using bi-encoder as reranker. Numbers in bold mean the result is the best excluding oracle. $\dag$ and $\ddag$ means $p < 0.1$ and $p < 0.05$ statistical significance over the Q0A baseline.}
\begin{tabular}{l|l|l|l|l|l|l}
\hline\hline
Users   & \multicolumn{3}{c|}{0-tolerance} & \multicolumn{3}{c}{1-tolerance} \\ \hline
Models  & R@1/100    & MRR    & Dec. err    & R@1/100    & MRR    & Dec. err     \\ \hline
Q0A     & 0.6725 & 0.7981   &   0.2425   & 0.6725   & 0.7981    & 0.2425     \\ 
Q1A     & 0.7200   & 0.7784  & 0.1400   & \textbf{0.7700}  &  \textbf{0.8352}  &  0.0725     \\ 
Q2A     &  0.0075  &  0.0075  & 0.9925   & 0.0075  &   0.0075     &   0.9925       \\ 
CtxPred & 0.7200 & 0.7784  &  0.1350  &  0.7675  &  0.8327      &  0.0700          \\ \hline
Ours    & \textbf{0.7375}$^{\ddag}$ &  \textbf{0.7982}  & \textbf{0.1225}$^{\ddag}$  &  0.7600         &  0.8337      &  \textbf{0.0650}$^{\ddag}$          \\ \hline
Oracle  & 0.8200 & 0.8898  &  0     &  0.8600  & 0.9156 & 0    \\ \hline\hline
\end{tabular}
\label{biexperiment}
\end{table*}

We want to specifically emphasize the comparison between our risk-aware agent and the risk-unaware CtxPred baseline which first predicts whether to ask clarifying question or just answer the query based on context, and then let the system ask or answer. From Table~\ref{polyexperiment} we see that the CtxPred baseline make more decision errors than our risk-aware agent ($0.1275>0.0975$). The reason can be found in Figure~\ref{distribution}, there are many conversations where CtxPred baseline correctly identifies the need for asking clarifying question but fails to find good clarifying question. As a result, it returns the user a bad question which gets 0 reciprocal score. Our risk-aware agent can reduce the number of times asking clarifying questions in these conversations and choose to answer the question instead. Thus our agent gets $3\%$ less decision error and $5.6\%$ higher MRR relatively. By this comparison, we show that it is necessary to jointly consider all the conversation context, the best retrieved answer, and the best retrieved clarifying question to decide whether to respond user with an answer or a question.

When user become more tolerant for bad clarifying questions, We see that Q1A, Q2A, and the oracle each has some performance improvements, because these models have more chances to find good clarifying questions to ask now. Another interesting fact is that models with lower decision errors also get higher MRR score. This is also because the decision of asking clarifying questions is now less likely to cause the user to leave and more likely to get user's response as additional information. Thus there are more conversations where the models get higher answer Reciprocal rank score, and there are less Reciprocal score loss due to user leaving. The most important observation is, as the user tolerance increases, we can see that our model's performance is converging to the Q1A baseline. This result is in our expectation since when user tolerance for bad clarifying questions increases, the risk in asking clarifying question will decrease. In the extreme case when user have infinite tolerance, the agent can ask any amount of bad clarifying questions until they ask a good one, then there will be no downside for asking clarifying question and thus there will be no risk for our risk-aware agent to be aware of. 

Our second set of experiments is to use bi-encoder as the reranker and test if our risk-aware model can work with a different reranker. We keep all experiment settings the same in the poly-encoder and bi-encoder experiments. The only difference is the reranker themselves. All the results for bi-encoder experiments are shown in Table~\ref{biexperiment}. From the table, we can see that all the model performances are slightly worse than them in poly-encoder because of the structure of the latter is more complex as it allows co-attention computation between the query context and candidate. In most experiments, our agent is able to outperform baselines in terms of answer quality and user experience. But when user tolerance becomes larger, the agent does not perform as well> We believe the reason is that the agent seeks a balance between Q0A and Q1A baselines, when Q0A is significantly worse than Q1A, our agent performance is also affected. Also, our model parameters is mainly tuned for poly-encoder structure, thus the bi-encoder experiments results can still be improved in the future. 


\begin{table*}[ht]
\caption{Ablation study results using poly-encoder as reranker. Numbers in bold mean the result is the best excluding oracle. $\dag$ and $\ddag$ indicates $p < 0.1$ and $p<0.01$ statistical significance over the best of encoded text and score models.}
\begin{tabular}{l|l|l|l|l|l|l|l|l|l}
\hline\hline
Users   & \multicolumn{3}{c|}{0-tolerance} & \multicolumn{3}{c|}{1-tolerance} & \multicolumn{3}{c}{2-tolerance} \\ \hline
Models  & R@1/100     & MRR    & Dec. err    & R@1/100     & MRR    & Dec. err    & R@1/100     & MRR    & Dec. err  \\ \hline
Encoded text     &  0.3919   & 0.4882    & 0.3236   & 0.4378 & 0.5284 & 0.2056  & 0.4611  & 0.5594    & 0.1631    \\ 
Score     &  0.4419  & 0.5428  & 0.2456   &  0.4519 &  0.5517  & 0.2350          &  0.4483 & 0.5505  & 0.2261  \\ 
Encoded text + Score  & \textbf{0.4461}    & \textbf{0.5435}   & \textbf{0.2375}$^{\dag}$  &   \textbf{0.4656}$^{\ddag}$ &  \textbf{0.5620}$^{\ddag}$ &   \textbf{0.1830}$^{\ddag}$  & \textbf{0.4788}$^{\ddag}$  & \textbf{0.5781}$^{\ddag}$ &  \textbf{0.1519}$^{\ddag}$         \\ \hline\hline
\end{tabular}
\label{ablation}
\end{table*}

\begin{figure}[t]
  \centering
  \includegraphics[width=\linewidth]{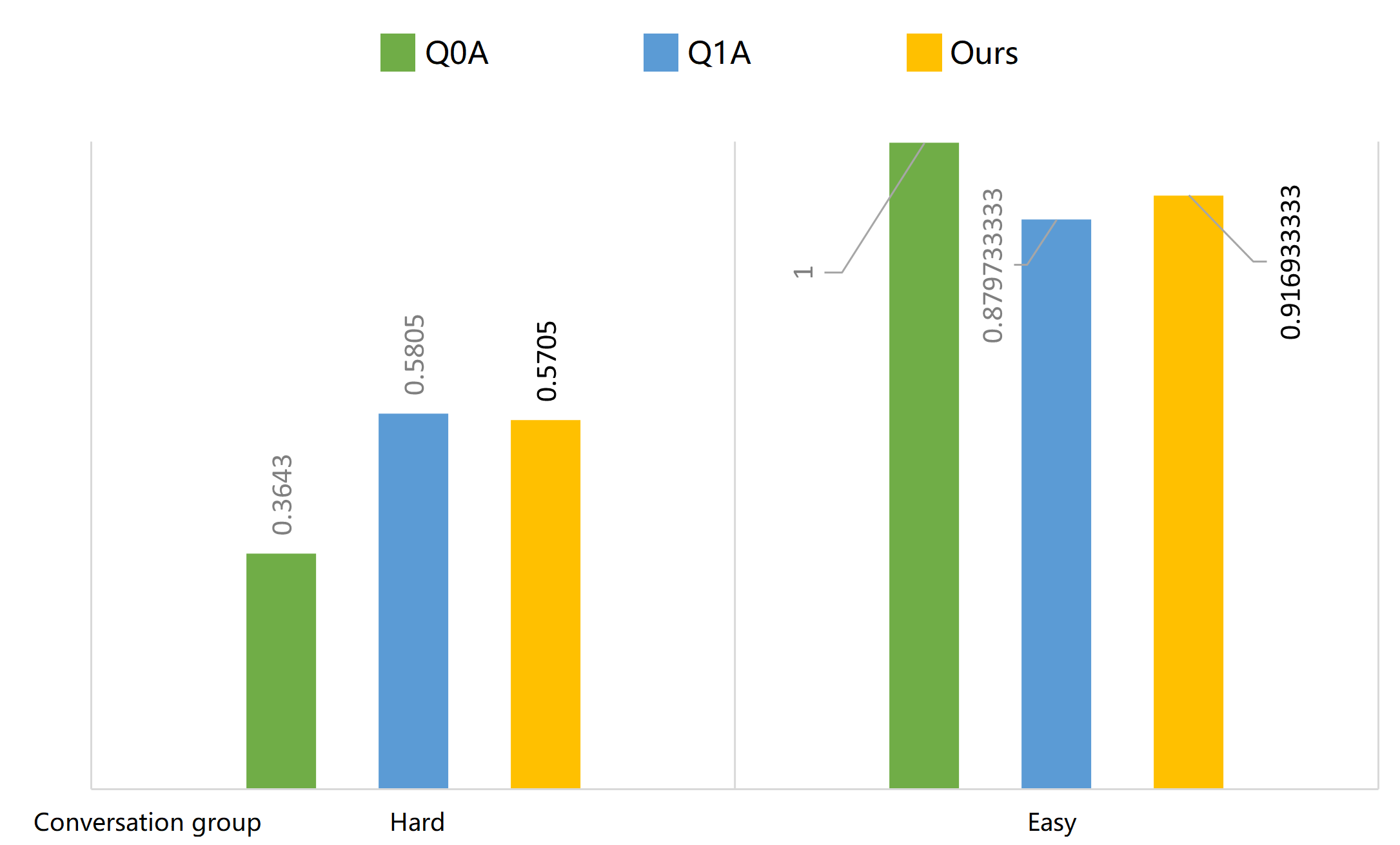}
  \caption{MRR Distributions in poly-reranker experiment with 0-tolerance user on 2 conversation groups.}
  \Description{MRR Distributions in poly-reranker experiment with 0-tolerance user on 2 conversation group.}
  \label{2bins}
\end{figure}

To get more insights of how our model improve answer quality in terms of MRR score. We split the test conversations of poly-encoder/0-tolerance user experiment into 2 conversation groups. The first group consists of hard conversations where Q0A baseline cannot directly answer, and the other group with easy conversations where Q0A can answer correctly. We compare the performances of Q0A and Q1A baselines, and our model for each group. The result in shown in Figure~\ref{2bins}. In the figure, the first group has the low overall MRR across all models, since it consists of the hardest conversations for Q0A and is generally hard for all models. Q1A improves Q0A on the first group by $0.2162$ MRR, but it also loses $0.1203$ MRR on the easier group. Consequently, the total MRR of Q1A is much less than Q0A. This result shows that if we allow a model to ask clarifying question freely to a intolerant user, the overall answer quality will be worse. The improvement of our model over Q0A on the first group is almost as much as Q1A ($0.2062$ versus $0.2162$). But on the easier group, our model is able to reduce the MRR loss from $0.1203$ to $0.0831$. From this comparison, we show that the reason of our model being better than the two baselines is that it is able to ask many good clarifying questions while minimizing the chance of asking bad ones. 

We further split the first hard conversation group into 2 smaller groups. The first group is made of conversations where Q0A always ranks the correct answer in second best, thus MRR = $0.5$. The other one are the rest even harder conversations where Q0A's MRR = $0.2386$. We compare our models' performances on this two groups again to see if our model's improvement is from the first easier group or the second harder ones. On the easier group, our model improves the MRR from $0.5$ to $0.7200$, on the harder bin, our model improves the MRR from $0.2386$ to $0.4310$. The improvement on easier group is slightly higher ($0.22>0.1924$). Considering the difference is not significant, we conclude that our model can help find better answer in all the conversations in general. 

In conclusion, our experiments compare our risk-aware agent with multiple baselines under different settings of reranker and user models. Our model is able to improve answer quality as well as user experience over all the baseline models. And the advantage of our model scales up when asking clarifying questions become harder and riskier.

\section{Ablation Study}

Our decision making deep Q network has a long feature list, consisting of mainly two type of features. The first is all the encoded text features for the query, context and candidates. The second is the ranking scores from reranker. We separate these two type of features and study their effects on our decision making DQN through ablation study. In ablation study, we have three decision making DQNs and their difference is input features. (1) Encoded text model uses only the encoded query $q$, context $h$, clarifying question candidates $\{cq_1,...cq_k\}$, and answer candidates $\{a_1,...a_k\}$. (2) Score model only uses the ranking scores $s_{cq}^{1:k}, s_{a}^{1:k}$ from poly-rerankers output. (3) Encoded text + Score model uses all the above features and is the model we tested in the main experiments.

The ablation study result is in Table~\ref{ablation}, using the same abbreviations as mentioned in result section. From the table we can see that both Encoded text model and score model perform worse than using all the features together ($0.7425<0.7550<0.7775$, etc.). This implies that all of the features actually play important roles in our decision module. The rankings scores features, despite small in amount, are indicative for evaluating the decision risks. We believe their usefulness should owe to the pretraining of poly-encoders on large Reddit dataset which leverages the transfer learning and multi-task learning effects. Our ablation study also shows that the decision module design and the reinforcement learning of the model is non-trivial since with only using the scores from pretrained rerankers cannot make good decisions.

\section{Conclusion}
In this paper, we bring to table the risk in conversational search, especially the risk of asking clarifying questions to user in conversational search. We show that although existing works study the benefits of asking clarifying questions and how to identify the need of asking clarifying questions, they neglect the fact that asking clarifying question is also risky and should not be taken as a default alternative to giving answer. To control and balance such risks, we propose a risk-aware conversational agent which make decisions between asking clarifying question and answering user's query by comprehensively evaluating and comparing the two actions. On training the agent, we also propose to use reinforcement learning to train the agent without having annotated data for when to ask clarifying question and when to give answer to the user. The original conversations are sufficient for training the agent. Through simulation experiments with different user models, we show that our risk-aware conversational search agent could improve both answer quality and user experience in interacting with the retrieval system.


\begin{acks}
This work was supported in part by the School of Computing, University of Utah. Any opinions, findings and conclusions or recommendations expressed in this material are those of the authors and do not necessarily reflect those of the sponsor.
\end{acks}


\bibliographystyle{ACM-Reference-Format}
\bibliography{sample-sigconf}

\appendix

\end{document}